\documentstyle[12pt]{article}
\begin{document}

\title{An Analysis of Wave Tails based on the Geometric Optics Approximation}
\author{Brien C. Nolan$^{a)}$\\
School of Mathematical Sciences,\\
Dublin City University,\\ Glasnevin, Dublin 9,\\
Ireland.
}
\maketitle
\begin{abstract}
The effect of the existence of tails on the propagation of scalar waves in 
curved space-time is considered via an analysis of flux integrals of the 
energy-stress-momentum tensor of the waves. 
The geometric optics approximation is formulated in terms of such flux integrals,
and three examples are investigated in detail in order to determine the possible
effects of wave tails. The approximation is valid for waves in Minkowski 
space-time (tail-free) and waves in Schwarzschild space-time (weak tails)
but it is shown how the approximation can break down in a cosmological scenario
due to destructive interference by strong tails. In this last situation, the 
waves do not radiate.
\end{abstract}
\newpage

\newcommand\newc{\newcommand}
\newc\esm{energy-stress-momentum\,\,}
\newc\be{\begin{eqnarray}}
\newc\ee{\end{eqnarray}}
\newc\del{\nabla}
\newc\fl[2]{\left.^{(#1)}\!F_{#2}\right.}
\newc\et[1]{\left._{{\rm{#1}}}\!T^{ab}\right.}
\newc\nn{\nonumber}
\newc\dop{{\dot\Psi}}
\newc\btu{\bigtriangleup}
\newc\xp{x^\prime}
\section{Introduction}
It is well known that scalar, electromagnetic and gravitational fields in curved 
space-times do not, in general, propagate entirely along the null cone, but are 
accompanied by `tails' which propagate in the interior of the null cone.
Indeed tail-free propagation appears to be the exception rather than the rule
\cite{frie,noon}. 
Wave tails manifest themselves in two ways; the ring down of a pulse of radiation
from a bound source \cite{price,gomez}, and partial back-scattering of the waves by the
curvature \cite{sonego1,press}. This latter effect may be represented graphically as 
follows (fig.1). A ray $\Gamma$ entering a curved region $\Omega$ of space-time is partially
transmitted ($\Gamma_T$) and partially reflected ($\Gamma_R$), or back-scattered,
by the curvature. 
It has been argued that in astrophysical situations, where
the curvature dies off rapidly away from the source, the existence of the
necessarily weak tails has a negligible effect on the propagation of 
radiation; the reflected portion $\Gamma_R$ diminishes rapidly far from the source 
\cite{thorne}. 
On the other hand, in cosmology, strong local curvature causes back-scatter
throughout the space-time, and tail effects may be considerable \cite{ellis}.
In fact it was shown in \cite{sonego1} that in the 
$k=-1$ Friedmann--Lema{\^\i}tre--Robertson--Walker
(FLRW) space--time, the usual definition of reflection and transmission coefficients 
breaks down. 

\begin{center}
Fig. 1\\
A simple representation of back-scattering by curvature.
\end{center}
\newpage

The aim of this paper is to discuss tail effects from the point of view of the
\esm tensor of the (scalar, electromagnetic, gravitational) perturbation.
The back-scattering effect is analysed by examining
the flux of the \esm of the field across certain fundamental surfaces.
Such scattering can be simply due to `potential barriers' in the equations which exist in the absence
of tails and are not curvature induced. Care is needed when seeking the source of 
reflection.

In particular, the geometric optics approximation for the field is considered.
In the usual approaches \cite{ehlers,anile,mtw}, this involves writing the field as 
an asymptotic series in a small parameter $\epsilon$ (short wavelength or inverse high frequency)
and analysing such high frequency modes. The results are well known; the light
rays are null geodesics, and to zeroth order in $\epsilon$, 
the field has an \esm tensor in the form of a null
fluid. The approach taken here is to look for a geometric optics approximation to  
the full physical field, rather than its Fourier transform, at a finite 
distance from the source.

Three representative cases are analysed in detail: minimally coupled, massless
scalar waves in Minkowski space-time, Schwarzschild space-time and the 
anti-Einstein static universe (the static $k=-1$ FLRW space-time). Scalar waves
are used for simplicity, similar results would be obtained for electromagnetic
and gravitational waves. The \esm of the field is most easily discussed in such
static space-times, where one has a locally conserved 4-momentum for the field,
leading to global conservation laws.

In \S 2, the fluxes of 4-momentum to be discussed are defined, and the geometric
optics approximation is described in terms of these.

In \S 3, these fluxes are calculated for a general retarded solution of the scalar
wave equation in Minkowski space-time, and the geometric optics approximation
is obtained. The same is done in \S 4 for Schwarzschild space-time, and the 
results are compared and contrasted with those of \S3. 

The effect of strong local curvature is seen in \S 5, where scalar fields in the 
anti-Einstein universe are examined. A Green's function approach is used, which
shows transparently the r\^ole played by the tail term. It is argued that the strong 
tail completely back-scatters the radiation field, leaving only non-radiative field
which does not escape to infinity.

The metric signature of $+2$ is used as are units in which $c=G=1$, and 
lower case Latin indices run from $0-3$.

\section{Flux Integrals and the Geometric Optics Approximation}

The Klein--Gordon equation is
\be \Box\Psi-\xi R\Psi-m^2\Psi& =& 0\,,\ee
where $\Box=g^{ab}\del_a\del_b$ and $R$ is the Ricci scalar of the space-time
in question, $\xi$ is a constant and $m$ is the mass of the scalar field $\Psi$.
In order that purely curvature induced effects are examined, only massless fields 
($m=0$) will be considered. In addition, for the three examples analysed in detail,
$\xi R=0$. Then the scalar field has a conserved symmetric \esm tensor,
\be T^{ab}&=&\del^a\Psi\del^b\Psi-{1\over2}g^{ab}\del_c\Psi\del^c\Psi\,,\label{esm}\ee
obeying $\del_a T^{ab}=0$. In a static space-time, one may then construct a
conserved 4-momentum for the field,
\be P^a&=&-T^{ab}\xi_b\,,\label{pdef}\ee
where $\xi^a$ is the time-like Killing field of the space-time. Then $P^a$ obeys
$\del_a P^a=0$.

Any static spherically symmetric space-time comes equipped with future null cones, 
$u=$ constant, where $u$ is retarded time, and surfaces of spherical symmetry,
$r=$ constant. Consider the 4-volume ${\cal{V}}$ bounded by the time-like 
surfaces $r=r_1$ and $r=r_2>r_1$, and by the future null cones $u=u_1$ and
$u=u_2>u_1$. Take $r_1$ large enough so that  ${\cal{V}}$ lies outside the
source of $\Psi$ (and outside the event horizon in the case of Schwarzschild).
The flux of $P^a$ across the surface $r=r_A (A=1,2)$ is
\be ^{(A)}\!F_T&=&
\int_{u_1}^{u_2} \int_{{\cal {S}}_2} P^a r_{,a}\sqrt{-g}\, d_3x\,,
\label{ft}\ee
where the $(A)$ indicates evaluation of the integral at $r=r_A$. 
The flux of
$P^a$ across the surface $u=u_A$ is
\be ^{(A)}\!F_R&=&
\int_{r_1}^{r_2} \int_{{\cal {S}}_2} P^a u_{,a}\sqrt{-g}\, d_3x\,,
\label{fr}\ee
where the $(A)$ indicates evaluation of the integral at $u=u_A$. 
The subscripts $T$ and $R$ indicate `transmitted' and `reflected'
respectively.
Integrating $\del_a P^a=0$ over ${\cal{V}}$ and using the divergence theorem 
shows that
\be \fl{1}{T}+\fl{1}{R}&=&\fl{2}{T}+\fl{2}{R}\,,\label{fluxcon}\ee
i.e. the flux of $P^a$ entering ${\cal{V}}$ via the inner and lower surfaces equals the
flux of $P^a$ leaving ${\cal{V}}$ via the outer and upper surfaces. 
See \cite{synge}
for more detail on such flux integrals in Minkowski space-time.

The generators of the future null cones have future-pointing tangent vector $k^a$,
where 
\be k_a&=&-u_{,a}\,,\ee and this will play the r\^ole of the ray vector for the field
$\Psi$. A geometric optics \esm tensor $\et{GO}$ is defined to be one for which
for any volume ${\cal{V}}$ defined above,
\begin{enumerate}
\item[1] $_{\rm{GO}}F_T$ is independent of $r$, which will imply that 
$\left. _{\rm{GO}}F_T\right|_{{\cal{J}}^+}$ is finite, and
\item[2] $_{\rm{GO}}F_R=0$.
\end{enumerate}
(Clearly no left-hand superscript is needed in either case to indicate through 
which surface of ${\cal{V}}$ the flux is calculated.)
The first condition says that $\Psi$ contributes a non-zero finite amount
of radiation at infinity (more precisely, at ${\cal{J}}^+$, where such is defined). 
The second condition indicates that there is no leakage of energy-momentum across
the null cones $u=$ constant; $_{\rm{GO}}\!P^a$ propagates entirely along the null rays
$k^a$. 
Such an \esm tensor
$\et{GO}$ satisfies Teitelboim's criterion for `radiation at arbitrary distances'
\cite{teit}. A geometric optics \esm tensor could also be described as a pure
radiation \esm tensor.

Condition (2) above is equivalent to $_{\rm{GO}}\!P^ak_a=0$, which in turn leads to 
$_{\rm{GO}}\!P^a=F(x)k^a$, since $_{\rm{GO}}\!P^a$ must be time-like or null. 
We continue
the definition of a geometric optics \esm tensor as folllows. The 4-momentum of
$\Psi$ as seen by an observer ${\cal{O}}$ with geodesic tangent 4-velocity $v^a$ is
\be P^a(v)&=&-T^{ab}v_b\,.\ee Demanding that {\em every} observer measures this to
be entirely in the outward null direction yields
\be T^{ab}v_b&=&G(x;v)k^a\,,\ee
which along with the conditions above gives the usual form for a geometric optics \esm tensor,
\be \et{GO}&=&H(x)k^ak^b\,.\ee

The geometric optics approximation is defined as follows \cite{he1}. 

{\em The physical \esm
tensor decomposes into two parts, $\et{RAD}$ (radiative) and $\et{REM}$ 
(remainder) which are separately 
conserved. $\et{RAD}$  is a geometric optics \esm tensor as defined above, and 
includes the} entire {\em radiation field, in the sense that $\et{REM}$ does
not contribute any radiation at ${\cal{J}}^+$:
\be \left._{\rm{REM}}\!F_T\right|_{{\cal{J}}^+} = 0\,.\ee
}
Demanding that $\et{RAD}$ and $\et{REM}$ be separately conserved is equivalent
to saying that the \esm of the field splits into two parts which propagate
independently of one another. Each piece will have associated with it a
conservation equation of the form (\ref{fluxcon}).
Intuitively, $\et{RAD}$ is the leading order term of $T^{ab}$ in an expansion
in inverse powers of some suitably defined distance parameter. Indeed, by hypothesis, 
when the geometric optics approximation holds, one has
\be \et{RAD}&=&H(x)k^ak^b\,,\label{emeqn}\ee
which is conserved. This leads to the geodesic equation for $k^a$, and (on 
transforming to an affine parameter along the rays if necessary) to
\be DH+\theta H&=&0\,,\label{fall}\ee 
where $D=k^a\del_a$ and $\theta=\del_a k^a$, 
which is an equation for the fall-off of $H$ in terms of the affine parameter.
This is a useful result, as it allows one to determine if a given \esm tensor includes
radiation, simply by looking for the leading order behaviour implied by (\ref{fall}).
If this strongest term is not present, the field does not radiate. This may be
demonstrated as follows.

The line-element for static spherical symmetry may be written as
\be ds^2&=&-f(r)du^2-2g(r)dudr+r^2d\omega^2\,,\label{ssle}\ee
where $d\omega^2$ is the line-element for the unit 2-sphere, $f$ and $g$ are positive
outside horizons, and $u$ is the retarded time. A convenient null tetrad is  
provided by $\{k^a,n^a,m^a,{\bar m}^a\}$, where
\be k_a&=&-u_{,a},\quad \Rightarrow k^a=g^{-1}\delta^a_1\,,\\
n^a&=&\delta^a_0-{1\over 2}fg^{-1}\delta^a_1\,,\quad m^a={1\over\sqrt{2}r}
(\delta^a_2+{i\over\sin\theta}\delta^a_3)\,.\ee
The only non-zero inner products are $k^an_a=-1$, $m^a{\bar m}_a=1$, and
$g_{ab}=-2k_{(a}n_{b)}+2m_{(a}{\bar m}_{b)}$. Then $\theta = 2/(rg)$,
and (\ref{fall}) may be integrated to obtain
\be H(x)&=&{H_0\over r^2}\,,\label{fo}\ee
where $H_0$ is independent of $r$. The time-like Killing field of (\ref{ssle})
is \be \xi^a&=&{1\over 2}fk^a+n^a\,,\ee
and so from (\ref{pdef}) and (\ref{emeqn})
\be _{\rm{RAD}}\!P^a&=&{H_0\over r^2}k^a\,.\ee
From this, one obtains
\be F_T&=& \int_{u_1}^{u_2}du\int_{{\cal{S}}_2} H_0\,d\omega\,,\ee
which is independent of $r$ and will be non-zero in general, leading to a non-zero
flux of radiation at ${\cal{J}}^+$. If $H(x)$ is any weaker than (\ref{fo}),
i.e. if $H(x)=o(r^{-2})$, then $F_T$ will be zero in the limit $r\rightarrow\infty$,
indicating that there is no radiation at ${\cal{J}}^+$.

The fluxes defined above are calculated for three representative examples in the
following sections, and it is shown how the presence of tails can influence
the existence of the geometric optics approximation defined here.

\section{Waves in Minkowski Space-Time}
In Minkowski space-time, the general solution of $\Box\Psi=0$ which
dies off as $r\rightarrow \infty$ and includes outgoing radiation
can be obtained as follows. Expanding $\Psi$ in spherical harmonics
$Y_{l,m}$, one
can write
\be\Psi=\sum_{l=0}^{\infty}\sum_{m=-l}^{+l}
\Psi^{(l,m)}Y_{l,m}\,.\label{m1}\ee 
The equation obeyed by $\Psi^{(l,m)}$ is (with 
indices suppressed)
\be 2{\dot \Psi}^\prime-\Psi^{\prime\prime}+{2\over r}({\dot\Psi}-\Psi^\prime)
+{l(l+1)\over r^2}\Psi&=&0\,,\label{lmeq}\ee
where a dot indicates partial differentiation with respect to $u$ and a prime
indicates partial differentiation with respect to $r$. 
For waves that die off as $r\rightarrow \infty$ one can write
\be\Psi=\sum_{n=1}^\infty \Psi^{(l,m)}_n(u)r^{-n}\,.\label{m2}\ee 
Filling out (\ref{lmeq}) and using static initial conditions one obtains
\be \Psi^{(l,m)}_n&=&{2^{l-n+1}l!(l+n-1)!\over (2l)!(n-1)!(l-n+1)!}
{d^{l-n+1}A_m(u)\over du^{l-n+1}}\,,\quad1\leq n\leq l+1\,,\nn\\
\Psi^{(l,m)}_n&=&0\,,\qquad n>l+1\,.\label{m3}\ee
The $A_m(u)$ are arbitrary functions of the null coordinate $u$. Thus the general
solution obtained from (\ref{m1}), (\ref{m2}) and (\ref{m3}) obeys 
$\Psi=O(r^{-1})$,
${\dot\Psi}=O(r^{-1})$, $\Psi^\prime=O(r^{-2})$ and $\delta\Psi=O(r^{-2})$ 
as $r\rightarrow\infty$, where $\delta=m^a\del_a$. Each of these relations
may be differentiated with respect to $r$.

A straightforward calculation yields
\be T^{ab}&=&({1\over2}\Psi^\prime-\dop)^2k^ak^b +(\Psi^\prime)^2n^an^b\nn\\
&&+2{\bar\delta}\Psi({1\over2}\Psi^\prime-\dop)k^{(a}m^{b)}
-2{\bar\delta}\Psi\Psi^\prime n^{(a}m^{b)} 
+({\bar\delta}\Psi)^2m^am^b + {\rm c.c.}\ee
The leading order term is easily identified;
\be \et{RAD}&=&{H_0\over r^2}k^ak^b\,,\label{m4}\ee
where
\be H_0(u,\theta,\phi)&=&\left(\sum_{l,m}{\dot\Psi}^{(l,m)}_1 Y_{l,m}\right)^2\,.\ee
This has the expected fall-off in terms of $r$, and is conserved. Thus so too
is $\et{REM}=T^{ab}-\et{RAD}$. The existence of $\et{RAD}$ in (\ref{m4}) above 
shows the validity of the geometric optics approximation as defined in \S 2 in
this situation.  
One has $_{\rm{RAD}}\!F_R=0$ (no leakage of the radiation field across
null cones) and $_{\rm{REM}}F_T=O(r^{-2})$ (no contribution at ${\cal{J}}^+$ from the
remainder field). 
It is worth pointing out that $_{\rm{REM}}\!F_R\neq0$, so that even in the absence of
curvature, some of the field is reflected or back-scattered.

\section{Waves in Schwarzschild Space-Time}
Scalar waves in Schwarzschild space-time have tails. In this section, it is shown
that the radiation field is unaffected by the existence of these tails, that
is, the geometric optics approximation remains valid, in the sense of \S 2. 
The starting point is Bardeen and Press's careful analysis of radiation fields
in the Schwarzschild background \cite{press}. These authors obtained the 
general outgoing radiation solution of the scalar wave equation in the form
\be \Psi&=&\sum_{l,m}\Psi^{(l,m)}(u,r)Y_{l,m}\,,\ee
where the coefficients (with indices
suppressed) are of the form
\be \Psi&=&\Psi^{{\rm I}}+\Psi^{{\rm II}}\,,\ee 
with
\be \Psi^{{\rm I}}&=&\sum_{n=1}^{l+1}f_n(u)r^{-n}\,,\\
\Psi^{{\rm II}}&=&r^{-l-1}\sum_{k=1}^\infty a_k({2m\over r})^k g_k(u,r)\,.\ee
The $g_k(u,r)$ are such that for each $l$, $\Psi^{{\rm II}}=O(r^{-l-2})$, and $\Psi^{(l,m)}$
is uniformly convergent for all values of the retarded time $u$.
As in the case of waves in Minkowski space-time, the properties of $\Psi^{{\rm I}}$
and $\Psi^{{\rm II}}$ above lead to $\Psi=O(r^{-1})$,
${\dot\Psi}=O(r^{-1})$, $\Psi^\prime=O(r^{-2})$ and $\delta\Psi=O(r^{-2})$ 
as $r\rightarrow\infty$, where $\delta=m^a\del_a$. Each of these relations
may be differentiated with respect to $r$. The \esm tensor of the field $\Psi$
is
\be T^{ab}&=&({1\over2}\btu\Psi^\prime-\dop)^2k^ak^b +(\Psi^\prime)^2n^an^b\nn\\
&&+2{\bar\delta}\Psi({1\over2}\btu\Psi^\prime-\dop)k^{(a}m^{b)} 
-2{\bar\delta}\Psi\Psi^\prime n^{(a}m^{b)} 
+({\bar\delta}\Psi)^2m^am^b + {\rm c.c.}\,,\label{semt}\ee
where $\btu=1-2m/r$ and $m$ is the usual Schwarzschild mass parameter.
Writing
\be\Psi&=&\Psi_0(u,\theta,\phi)r^{-1}+O(r^{-2})\,,\ee
the leading order term in (\ref{semt}) is found to be
\be \et{RAD}&=&{\dop_0^2\over r^2}k^ak^b\,,\ee
which has the predicted fall-off and is conserved. The remainder term $\et{REM}=T^{ab}-\et{RAD}$ is then also 
conserved. Again as in Minkowski space-time, the geometric optics approximation is 
valid, and
one has $_{\rm{RAD}}\!F_R=0$ (no leakage of the radiation field across
null cones) and $_{\rm{REM}}F_T=O(r^{-2})$ (no contribution at ${\cal{J}}^+$
from the remainder field). Thus
there is no qualitative difference between the propagation of the {\em radiation}
field in Minkowski and Schwarzschild space-time. Genuine back-scattering of
other parts of the field was detected by Bardeen and Press\cite{press}, who showed that
$\Psi^{{\rm II}}$ is constructed from terms which can be written in the form of
ingoing waves on a flat background.

\section{Waves in the anti-Einstein Universe}
To illustrate the extreme properties that wave tails may have, consider the
wave equation 
\be (\Box+\xi R)\Psi&=&j\label{we}\ee
on the static $k=-1$ FLRW space-time (the anti-Einstein universe), for which
the line element is
\be ds^2&=&-dt^2+dr^2+\sinh^2rd\omega^2\,.\label{le}\ee
In (\ref{we}), $j$ is a source term and $\xi$ is a constant.

The leading order term in the \esm tensor of $\Psi$ should be the
geometric optics term,
\be \et{RAD}&=&H(x)k^ak^b\,,\label{cosesmt}\ee
where $k^a=-g^{ab}u_{,b}$ is the outgoing null geodesic direction. Conservation of
(\ref{cosesmt}) and properties of $k^a$ deduced from (\ref{le}) show that
\be H(x)&=&{H_0(u,\theta,\phi)\over\sinh^2r}\,.\label{flof}\ee
Thus in order that $\et{RAD}$ may be constructed, one must have
\be \Psi(x)&=&{\Psi_0(u,\theta,\phi)\over \sinh r} + o({1\over\sinh r})\,.\ee
The object of the rest of this section is to show that this condition is {\em not}
satisfied by quite general solutions $\Psi$ of the minimally coupled wave equation.

Following the Green's
function method of solution of DeWitt and Brehme \cite{dewb}, Hadamard's 
elementary solution of the homogeneous version of (\ref{we}) is
\be G^{(1)}(x,x^\prime)&=&{1\over 4\pi}(\Sigma\Omega^{-1}+V\ln\Omega+W)\,,\label{g1}\ee
where $\Sigma,V$ and $W$ are two-point functions on space-time, free of singularities
and $\Sigma$ obeys
\be [\Sigma]&\equiv&\lim_{x\rightarrow x^\prime}\Sigma(x,x^\prime)=1\,.\ee
$\Omega$ is the world function \cite{sygr} of the space-time, given by, in the case of
(\ref{le}) above, 
\be \Omega(x,\xp)&\equiv&-{1\over 2}s^2=-{1\over 2}((t-t^\prime)^2-\rho^2)\,,\ee
where
\be \cosh\rho&=&\cosh r\cosh r^\prime-(\sin\theta\sin\theta^\prime\cos(\phi-\phi^\prime)+
\cos\theta\cos\theta^\prime)\sinh r\sinh r^\prime\,.\nn \\ \ee
See \cite{he} for more details. The retarded Green's function is constructed from
(\ref{g1}) and is given by
\be G^R(x,\xp)&=&{1\over 4\pi}(\Sigma\delta(\Omega)-V\theta(-\Omega))\theta(t-t^\prime)\,.\label{gr}\ee
Then the retarded solution of (\ref{we}) is
\be \Psi&=&\int G^R(x,\xp)j(\xp)\,d_4\xp\,.\label{retsol}\ee
For the space-time in question, one can show that 
\be \Sigma&=&{\rho\over\sinh\rho}\,,\ee
and the Hadamard series expansion for $V$ is
\be V&=&{\rho\over\sinh\rho}\sum_{n=0}^\infty 
{(1-6\xi)^{n+1}\over 2^{n+1}n!(n+1)!}\Omega^n\,.\label{hser}\ee
This is the tail term for the propagation of the field $\Psi$; (\ref{gr}) shows
that $V$ contributes in the interior of the null cone, where $\Omega<0$.
Concentrating on two representative examples shows clearly the r\^ole played 
by $V$. For conformal coupling, $\xi=1/6$ and $V=0$. This is as expected, the 
propagation of $\Psi$ on the conformally flat background of (\ref{le}) is
equivalent to propagation on Minkowski space-time. For minimal coupling, $\xi=0$ and
one can write
\be V&=&{\rho\over\sinh\rho}s^{-1}J_1(s)\,,\label{visbes}\ee
where $J_1$ is a Bessel function of the first kind.

To consider outgoing waves from a bound source, take $j(x)$ to be confined to
a time-like world-line $C: x=z(\tau)$ where $\tau$ is proper time along $C$
(see \S 5.6 of \cite{frie}). For a monopole source, which would dominate the radiation field,
one may write
\be j(\xp)&=&\int_{-\infty}^{+\infty} q(\tau)\delta^{(4)}(\xp,z(\tau))\,d\tau\,.
\label{source}\ee
Splitting $\Psi$ into the sharply propagated term and the tail term and integrating
over $\xp$ yields $4\pi\Psi=f-g$, with
\be f(x)&=&\int_{-\infty}^\infty 
\Sigma\delta(\Omega)\theta(t-z^0(\tau))q(\tau)\,d\tau\,,
\label{fsol}\\
g(x)&=&\int_{-\infty}^\infty 
V\theta(-\Omega)\theta(t-z^0(\tau))q(\tau)\,d\tau\,.
\label{gsol}\ee
The latter point in two point functions is now $\xp=z(\tau)$. For convenience, and
with no significant loss of generality, take $C$ to be geodesic, which one can then
assume to be $r=0$. Then $z^a=\tau\delta^a_0$ and $\rho=r$. The delta function
can be expanded as
\be \delta(\Omega)&=&{\delta(\tau-u)\over r}+{\delta(\tau-v)\over r}\,,\ee
where $u\equiv t-r$ is retarded time and $v\equiv t+r$ is advanced time. One can then
write
\be f(x)&=&{q(u)\over \sinh r}\,,\label{csol}\ee
and
\be g(x)&=&{r\over\sinh r}\int_{-\infty}^u s^{-1}J_1(s)q(\tau)\,d\tau\,,
\label{tsol}\ee
where now  $s=[(t-\tau)^2-r^2]^{1/2}\geq 0$. Eqn.(\ref{csol}) yields the solution
for the conformally coupled equation.

The behaviour of $g(x)$ as $r\rightarrow\infty$ is determined as follows. Write
$g(x)=I(x)\left\{r/\sinh r\right\}$. Integrating by parts once using
$d(J_0(s))=-s^{-1}J_1(s)(\tau-t)d\tau$ and asymptotic properties of $J_0$ 
\cite{abr}
yields
\[ I(x)={q(u)\over\sinh r}-I_1(x)\,,\]
so that 
\be \Psi(x)&=&{1\over4\pi}{r\over\sinh r}I_1\,,\label{521}\ee 
where
\be I_1&=&\int_{-\infty}^u J_0(s)\left\{
{q^\prime(\tau)\over (s^2+r^2)^{1/2}}+{q(\tau)\over s^2+r^2}\right\}\,d\tau\,.\ee
Next, assume that the source switched on at some finite time in the past,
which we take to be $\tau=0$. Then $q(\tau)=q^\prime(\tau)=0$ for $\tau<0$,
and $I_1=I_2+I_3$, with
\be I_2(x)&=&\int_0^u {J_0(s)\over (s^2+r^2)^{1/2} } q^\prime(\tau)\,d\tau\,.\ee
Using Schwarz's inequality, one may write
\be |I_2|^2&\leq&M(u)I_4\,,\ee
where
\[ M(u)=\int_0^u (q^\prime(\tau))^2\,d\tau\]
is positive and
\be I_4(x)&=&\int_0^u {J_0^2(s)\over s^2+r^2}\,d\tau =
\int_0^{s_1}{sJ_0^2(s)\over(s^2+r^2)^{3/2}}\,ds\,,\ee
where $s_1=(u^2+2ur)^{1/2}$. Integrating by parts using $sJ_0^2(s)ds=
d\left\{{s^2\over 2}(J_0^2+J_1^2)\right\}$ gives
\be I_4&=&{s_1^2\over 2(s_1^2+r^2)^{3/2}} (J_0^2(s_1)+J_1^2(s_1))+
{3\over 2}\int_0^{s_1} {s^3\over (s^2+r^2)^{5/2}} (J_0^2+J_1^2)\,ds\,.\ee
In the limit $r\rightarrow\infty$, $u$ finite (i.e. near ${\cal{J}}^+$), the
first term in $I_4$ is
\be {u^2+2ur\over 2(u+r)^3}\left\{{2\over \pi} (u^2+2ur)^{-1/2}+O(r^{-1})
\right\}&=&{(2u)^{1/2}\over \pi}r^{-5/2}+O(r^{-3})\,.\ee
Using the upper bound $J_0^2+J_1^2\leq 3/2$, the second term in $I_4$ is
dominated by
\[ {9\over 4}\int_0^{s_1} {s^3\over (s^2+r^2)^{5/2}} \,ds
={9\over 4}u^2r^{-3}+O(r^{-4})\,.\]
Thus one has the bound
\be |I_4|&\leq&{(2u)^{1/2}\over \pi}r^{-5/2} + O(r^{-3})\,,\ee
which gives $I_2=O(r^{-5/4})$.

A bound for $I_3$ is obtained more easily. Again using Schwarz's inequality,
\be |I_3|^2&\leq&N(u)I_5\,,\ee
where
\[ N(u)=\int_0^u q^2(\tau)\,d\tau\]
is positive and
\be I_5&=&\int_0^{s_1}{s\over(s^2+r^2)^{5/2}}J_0^2(s)\,ds\,.\ee
Using $|J_0|\leq 1$ gives
\be I_5&\leq&\int_0^{s_1}{s\over(s^2+r^2)^{5/2}}\,ds
=ur^{-4}+O(r^{-5})\,,\ee
so that $I_3=O(r^{-2})$.

The overall result is $I_1=O(r^{-5/4})$, and so
\be \Psi(x)&=&{1\over 4\pi\sinh r}O(r^{-1/4})\,.\label{cent}\ee
Thus $\Psi$ falls off more rapidly than the rate necessary for the field to radiate.
The separately conserved geometric optics \esm tensor $\et{RAD}$ cannot be
constructed, and in addition,
\be F_T|_{{\cal{J}}^+}&=&\lim_{r\rightarrow\infty} \int -T^{ab}\xi_ar_{,b}\, d_3v =0\,,
\ee
so that there is no radiation at ${\cal{J}}^+$.
The interpretation of this result is clear. The tail term $V$, which is strongest on
and near the null cone $\Omega=s=0$, backscatters the 
{\em radiation field itself},
rather than just other non-radiative components of the field.

The situation for the solution $\Psi(x)=Q(u)/\sinh r$ of the tail-free conformally invariant
equation (obtained from (\ref{csol})) is more familiar. The appropriate
\esm tensor is \cite{penrin2}
\be T^{ab}&=&\del^a\Psi\del^b\Psi-{1\over 4}g^{ab} \del_c\Psi\del^c\Psi
-{1\over 2}\Psi\del^a\del^b\Psi-{1\over 4}\Psi^2(R^{ab}-{1\over 6}Rg^{ab})\,,\ee
which for $\Psi$ given above yields
\be T^{ab}&=&\left\{{1\over \sinh^2r}
({\dot Q}^2-{1\over2}Q{\ddot Q}+{1\over 2}Q{\dot Q})
+{1\over 2}{e^{-r}\over\sinh^3r}Q{\dot Q}\right\}k^ak^b\nn\\
&&+{1\over 2}{Q^2\over\sinh^4r}(k^{(a}n^{b)}+m^{(a}{\bar m}^{b)})\,.\ee
The leading order term is readily identified as
\be\et{RAD}&=&{1\over\sinh^2r}
({\dot Q}^2-{1\over2}Q{\ddot Q}+{1\over 2}Q{\dot Q})k^ak^b\,,\label{cosrad}\ee
which displays the fall-off as $r\rightarrow\infty$ predicted in (\ref{flof})
above. As in \S\S 3 and 4 above, 
one has $_{\rm{RAD}}\!F_R=0$ (no leakage of the radiation field across
null cones) and ${\rm lim}_{r\rightarrow\infty} {}_{\rm{REM}}\!F_T=0$ (no contribution at ${\cal{J}}^+$ from the
remainder field). The energy radiated through any $r=$ constant surface in
the retarded time $u_2-u_1$ is
\be _{\rm{RAD}}F_T&=&4\pi\int_{u_1}^{u_2}
({\dot Q}^2-{1\over2}Q{\ddot Q}+{1\over 2}Q{\dot Q})\,du\,,\ee
which also gives the radiation rate at ${\cal{J}}^+$. The coefficient of $k^a$ 
in 
\[ _{\rm{RAD}}\!P^a={1\over\sinh^2r}
({\dot Q}^2-{1\over2}Q{\ddot Q}+{1\over 2}Q{\dot Q})k^a\]
should be positive to give a future pointing 4-momentum, which imposes restrictions
on the source term $Q(\tau)$.

\section{Discussion}
The central technical result of \S 5 is eq.(\ref{cent}), the `too weak to radiate'
condition. This behaviour is generic, and does not depend on the assumptions about the source
term $q(\tau)$ implicit throughout the calculation, nor indeed on the use of
a monopole solution. This can be seen as follows. 

The general solution for outgoing waves of the minimally coupled wave equation 
in the anti-Einstein universe can be written as \cite{sonego1}
\be \Psi(x)&=&{1\over\sinh r} \int_{-\infty}^{\infty} e^{-i\omega t}
\sum_{l,m} A_{l,m}(\omega) e^{ipr} F_{l,p}(z) Y_{l,m}\,d\omega\,,\label{sol}\ee
where $p=\sqrt{\omega^2-1}$, the $F_{l,p}$ are (hypergeometric) polynomials
\be F_{l,p}(z)&=&\sum_{n=0}^l \left( {{l+n}\atop l}\right)
{(-l)_n\over (1-ip)_n} z^n\,,\ee
$z=(1-e^{2r})^{-1}\in (-\infty,0)$ and the coefficients $A_{l,m}(\omega)$
arise from a time Fourier transform of a source term. Then using the 
Riemann-Lebesgue lemma and a minimal assumption on the behaviour of the source,
one has, for each $l,m$ \cite{roseau},
\be A_{l,m}(\omega)&=&o(\omega)\,,\quad \omega\rightarrow\pm\infty\,.
\label{rll}\ee
Thus to obtain the large $r$, finite $u$, behaviour of (\ref{sol}), one is lead to
consider integrals of the form
\be I(r,u)&=&\int_\Gamma A(\omega) e^{-i\omega u} e^{iqr}\,d\omega\,,\label{intform}\ee
where $q=\sqrt{\omega^2-1}-\omega$, $A(\omega)=o(\omega)$ as $\omega\rightarrow\pm\infty$
and apart from slight deformations, the contour $\Gamma$ is of infinite extent,
running from $Re(\omega)=-\infty$ to $Re(\omega)=+\infty$. Thus the integral 
is of the form examined in \S9.5 of \cite{bh}; the results obtained are as 
follows. See the given reference for more details.

The relevant limit of (\ref{intform}) is $r\rightarrow +\infty$, $u$ finite.
Then one may assume (for $u$ positive; $u$ negative leads to similar results)
\be \epsilon &\equiv&{u\over r}\in [0,\alpha)\,,\ee
where $\alpha\ll 1$. A {\em uniform} asymptotic expansion for (\ref{intform})
with $\epsilon$ in the given range is obtained in terms of Bessel functions 
of the first kind, with leading order behaviour
\be I(r;\epsilon)&\sim& b_0(\epsilon)I_0(r,\epsilon)
+b_1(\epsilon)I_1(r,\epsilon)\,,\label{asymp}\ee
where $b_n=O(1)$ as $\epsilon\rightarrow 0$, and
\be I_n&=&-2\pi i(2\gamma e^{-i\pi/2})^{\lambda+n} J_{\lambda+n}(\gamma r)\,,
\ee
with $\gamma=(\epsilon^2+\epsilon)^{1/2}$. The constant $\lambda$ will be 
strictly positive (as a consequence of (\ref{rll})), and for a source consisting of 
an oscillator at the origin, $\lambda =1$ \cite{bh}. Since the result (\ref{asymp})
is uniform in $\epsilon$, one can substitute $\epsilon=u/r$ and obtain
(using asymptotics of Bessel functions \cite{abr})
\be I(r,u)&=&O(r^{-1/4})\,.\ee
This is the behaviour in the limiting case $\lambda=0$; the more reasonable 
$\lambda=1$ yields
\be I(r,u)&=&O(r^{-3/4})\,.\ee
In any case, for very general sources, one finds
\be \Psi(x)&=&{1\over\sinh r}O(r^{-1/4})\,,\ee
which is eq.(\ref{cent}). The calculation given in \S 5 was preferred as it
shows transparently the r\^ole played by the tail term.

An important question is whether or not the behaviour of scalar waves in the
anti-Einstein universe is typical for waves in matter. Clearly, this will not
be the case for electromagnetic waves in an FLRW universe, which have no tails.
The interference of the tail with the radiation field arises from that part of
the Green's function tail term $V$ which is non-zero on the null cone. This
is integrated out in the calculation preceding eq.(\ref{521}), and cancels out the
sharply propagated term. Consider the tail term $V^{a^\prime}_b$ for the
electromagnetic Green's function. A one-point expansion may be performed to 
give \cite{dewb}
\be V^{a^\prime}_b(x,\xp)&=&g^{a^\prime}_b \sum_{n=0}^\infty 
{V^c}_{b{b_1}...{b_n}}(x) \del^{b_1}\Omega\cdots\del^{b_n}\Omega\,,
\label{1pexp}\ee
where $g^{a^\prime}_b$ is the parallel propagator of the space--time.
In general, $\del_a\Omega=O(s)$, and along the null cone, $\del_a\Omega(x,\xp)$ is
the null geodesic direction at $x$, pointing from the source point $\xp$ to 
the field point $x$.
The first term in (\ref{1pexp}) is
\be V_{ab}&=&R_{ab}-{1\over 6}Rg_{ab}\,,\ee
and the condition for this to be non-pure gauge is \cite{sonego2}
\be \del_{[c}R_{b]a}-{1\over 6} \del_{[c}Rg_{b]a}&\equiv&\del^d C_{bcad}\neq 0\,.\ee
In an almost-FLRW universe, this term can make small contributions to the tail
along the null cone, as could higher order terms in (\ref{1pexp}), which 
can begin the annihilation of the radiation field which is total in \S 5. This
may have implications for the traditional use of the geometric optics approximation
in cosmological observations \cite{ellis2}, and warrants a more thorough investigation.

The author suggests that the geometric optics approximation 
discussed above provides a convenient way of
determining if wave tails have a significant effect on the propagation of 
radiation in {\em any} space-time. The characterisation of a wave tail being
strong - failure of the geometric optics approximation as formulated above -
was motivated by analysing fluxes in space-times with nice global properties,
but is stated in purely local terms and may be investigated in any space-time 
with asymptotic regions. The interpretation of the approximation is clear;
there exists a part of the field (the radiation field) whose \esm tensor is
$\et{RAD}$, which detaches itself from the source and travels away to 
infinity at the speed of light, propagating independently of the rest of the field 
($\et{RAD}$ is separately conserved). As viewed by any observer, the radiation field
is in the outward null direction, and includes all of the radiation emitted by
the source of the field.
The strength of wave tails for Maxwell's equations
is currently being investigated in these terms in asymptotically flat space-times,
wherein it is fully expected that the geometric optics approximation is valid, 
and hence the tails are insignificant \cite{thorne}.

Finally, it is tempting to consider an extension of this discussion to the 
gravitational field. This has already been carried out in the case of the
linearized theory\cite{me}; it would be interesting to see the relationship with the
quasi-localisation of Bondi--Sachs mass loss \cite{hay}.

\section*{Acknowledgements}
I am indebted to Chris Luke for many useful conversations and for introducing me 
to asymptotics.



\end{document}